\documentclass[pra,twocolumn,showpacs,preprintnumbers,amsmath,amssymb]{revtex4-1}
\usepackage{graphicx}
\usepackage{dcolumn}
\usepackage{bm}
\usepackage{latexsym,epsfig}
\usepackage{textcomp}

\begin{document}
\title{Three body on-site interactions in ultracold bosonic atoms in an optical lattice and superlattice}

\author{Manpreet Singh$^{1,2}$}
\author{Arya Dhar$^{1}$}
\author{Tapan Mishra$^{3}$}
\author{R. V. Pai$^{4}$}
\author{B. P. Das$^{1}$}
\affiliation{$^{1}$ Indian Institute of Astrophysics, Bangalore 560 034, India}
\affiliation{$^{2}$ School of Interdisciplinary and Transdisciplinary Studies, IGNOU, New Delhi 110 068, India}
\affiliation{$^{3}$ International Centre for Theoretical Sciences (ICTS), Bangalore 560 012, India}
\affiliation{$^{4}$ Department of Physics, Goa University, Taleigao Plateau, Goa 403 206, India}

\date{\today}

\begin{abstract}

The Mott insulator-superfluid transition for ultracold bosonic atoms
in an optical lattice has been extensively studied in the framework
of the Bose-Hubbard model with two-body on-site interactions. In
this paper, we analyze the additional effect of the three-body
on-site interactions on this phase transition in an
optical lattice and the transitions between the various phases that
arise in an optical superlattice. Using the mean-field theory and
the density matrix renormalization group method, we find the phase
diagrams depicting the relationships between various physical
 quantities in an optical lattice and superlattice.
We also propose a possible experimental signature to observe the on-site three-body interactions.

\end{abstract}



\maketitle

Studies of ultracold bosonic atoms in optical lattices offer many
opportunities for exploring a variety of quantum phases. After the
first theoretical prediction of superfluid (SF) to Mott insulator
(MI) transition in bosonic systems by Fisher \textit{et al.}~\cite{fisher}, in
three dimensional optical lattice by Jaksch \textit{et al.}~\cite{jaksch}
which was followed by an experimental observation by Greiner et
al~\cite{greiner}, a number of quantum phases have been reported in
the literature~\cite{lewenstein, bloch}.
 The ability to fine tune the control parameters in such experiments makes it possible
to probe these exotic quantum phases ~\cite{ibloch, wketterle}.
 Apart from the hopping and the on-site two body interactions,
the on-site three-body interactions can also influence the onset of different phases.
Zhang \textit{et al.} had earlier found the extension of the insulating
lobes in the presence of the on-site three-body interactions,
using the decoupling mean-field theory~\cite{zhang}. The generation
of effective three- and higher-body interactions by two-body
collisions of atoms confined in the lowest vibrational
 states of a three-dimensional optical lattice has been reported by
Johnson \textit{et al.}~\cite{tiesinga}. The effect of three-body interactions on the insulating lobes
in an optical lattice has been considered using the mean-field and
functional integral approaches in the Bose-Hubbard approximation for optical
lattices~\cite{jaksch, zhou}. On the other hand, studies on optical superlattices
have revealed the existence of phases other than the usual MI and SF phases;
namely the superlattice induced Mott insulator(SLMI) phases which have a density
modulation in the system~\cite{aryadmrg, aryamft}.

On the experimental side, Will \textit{et al.}~\cite{sewill} have detected and
precisely measured the on-site three and higher body interaction
strengths by observing the collapse and revival of the superfluid
matter waves in a deep optical lattice. N\"{a}gerl et
al~\cite{nagerl} have been able to precisely determine the on-site
interaction energies including multi-body interaction shifts. In
another work, Greiner \textit{et al.}~\cite{ruichao} have determined the
three-body interaction strengths by using occupation-sensitive
photon-assisted tunneling. However, the effect of three-body
interactions in optical superlattices has still not been
investigated to the best of our knowledge.

In this paper, we study the effect of the on-site repulsive
three-body interactions in addition to the on-site two-body
interactions on various phases exhibited by ultracold bosonic atoms in an optical lattice and a
superlattice. We use the mean-field decoupling approximation and the
finite size density matrix renormalization group (FS-DMRG) method
for various densities and three-body interaction strengths and then
compare the results.
The system of bosons in an optical superlattice with three-body interaction can be described by the modified
Bose-Hubbard model as follows:-
\begin{eqnarray}
 H=-t\sum_{\langle{i,j}\rangle}{(\hat{a}_{i}^{\dagger}\hat{a}_{j}+\mbox{H.c})}
   +\frac{U}{2}\sum_{i}{\hat{n}_{i}(\hat{n}_{i}-1)} \nonumber\\
   +\frac{W}{6}\sum_{i}{\hat{n}_{i}(\hat{n}_{i}-1)}{(\hat{n}_{i}-2)}
   -\mu{\sum_{i}{\hat{n}_{i}}}+\sum_{i}{\lambda_{i}\hat{n}_{i}}
 \label{eq:one}
\end{eqnarray}
Here, $\hat{a}_{i}^{\dagger}$($\hat{a}_{i}$) is the creation
(annihilation) operator which creates (destroys) an atom at site
$i$, $\hat{n}_{i}=\hat{a}_{i}^{\dagger}\hat{a}_{i}$ is the number
operator, $t$ is the hopping amplitude between the adjacent sites
$\langle{i,j}\rangle$, $U$ and $W$ represent the on-site
inter-atomic two-body and three-body interactions respectively,
$\mu$ is the chemical potential, and $\lambda$ is the superlattice
potential. We consider
\begin{figure}[h]
\begin{center}
\includegraphics[height=4.4cm,width=6.8cm]{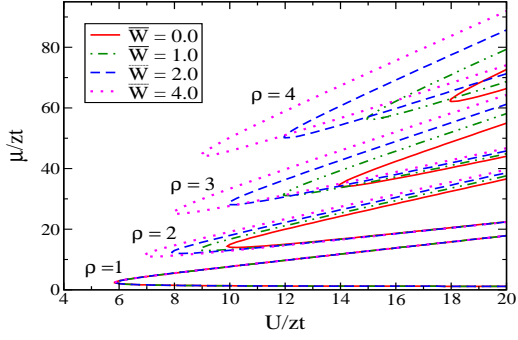}
\caption{(Color online) Phase diagram of Eq.~(\ref{eq:two}) for
different $\overline{W}$ for optical lattice. The lobes represent Mott
insulator phases for densities $\rho=1,2,3,4$.}
 \label{fig:fig1}
\end{center}
\end{figure}
 a bipartite lattice with sublattices ${\mathcal
A}$ and ${\mathcal B}$ with a periodicity of two sites.
We apply standard decoupling
approximation~\cite{sheshadri,stoof,aryamft} to the hopping term in
Eq.~(\ref{eq:one}) to obtain the mean-field Hamiltonian given by
\begin{eqnarray}
\frac{H^{MF}_{i}}{zt}&=&-\phi_{i}{(\hat{a}_{i}^{\dagger}+\hat{a}_{i})}+\phi_{i}\psi_i
   +\frac{\overline{U}}{2}{\hat{n}_{i}(\hat{n}_{i}-1)} \nonumber\\
   &&+\frac{\overline{W}}{6}{\hat{n}_{i}(\hat{n}_{i}-1)}{(\hat{n}_{i}-2)}
   -\overline{\mu}{\hat{n}_{i}}+\overline{\lambda}_i{\hat{n}_{i}}
 \label{eq:two}
\end{eqnarray}
where the superfluid order parameter $\psi_i=<\hat{a}_i>$ is taken
to be real~\cite{sheshadri},
$\phi_i=\frac{1}{z}\sum_{\delta}\psi_{i+\delta}$, the summation over
$\delta$ is taken over $z$ nearest neighboring sites,
$\overline{U}=U/zt$, $\overline{W}=W/zt$, $\overline{\mu}=\mu/zt$
and $\overline{\lambda}_i=\lambda_i/zt$ are dimensionless
parameters. For an optical lattice, $\lambda_i=0$  for all $i$, thus
$\psi_i=\psi$. For our optical superlattice, $\lambda_i=0$ for
sublattice ${\mathcal A}$ and $\lambda_i=\lambda$ for sublattice
${\mathcal B}$, thus $\psi_i=\psi_A (\psi_B)$ if $i$ belongs to
sublattice ${\mathcal A}$ (${\mathcal B}$). The mean-field eigen
value equation is solved self-consistently to obtain the local
superfluid density $\rho^s_i=\psi^2_i$ and density
$\rho_i=\langle\hat{n}_i\rangle$ of the ground state of the system.

\begin{figure}[b]
\psfig{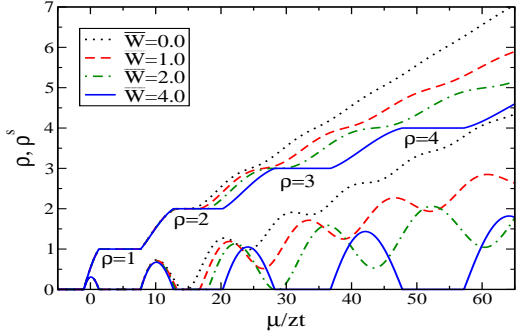}
\caption{(Color online) Variation of density $\rho$ and superfluid density $\rho^s$
with $\overline{\mu}$ for $\overline{U}=10$, for optical lattice. Top to bottom, the first
four curves represent density $\rho$ and the next four curves represent
superfluid density $\rho^s$. The plateaus in the $\rho$ plots represent
MI phases with vanishing $\rho^s$.}
 \label{fig:fig2}
\end{figure}

To study the effect of $\overline{W}$ on MI phases in an optical lattice,
we first present the mean-field phase diagram for an optical lattice (Fig.~\ref{fig:fig1}), 
in the $\overline{U}$ - $\overline{\mu}$ plane obtained from the density
$\rho$ and the superfluid density $\rho^s$, for various values of
$\overline{W}$. Figure~\ref{fig:fig2} shows the $\overline{\mu}$ -
$\rho,\rho^s$ plot for different $\overline{W}$.
\begin{figure}[t]
 \centering
\psfig{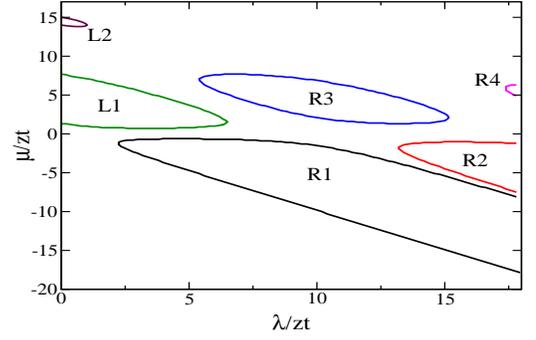}
\caption{(Color online) $\overline{\lambda}$ - $\overline{\mu}$ phase diagram
for $\overline{U}=10$, $\overline{W}=0.0$, for optical superlattice.}
 \label{fig:fig3}
\end{figure}
\begin{figure}[b]
 \centering
\psfig{file=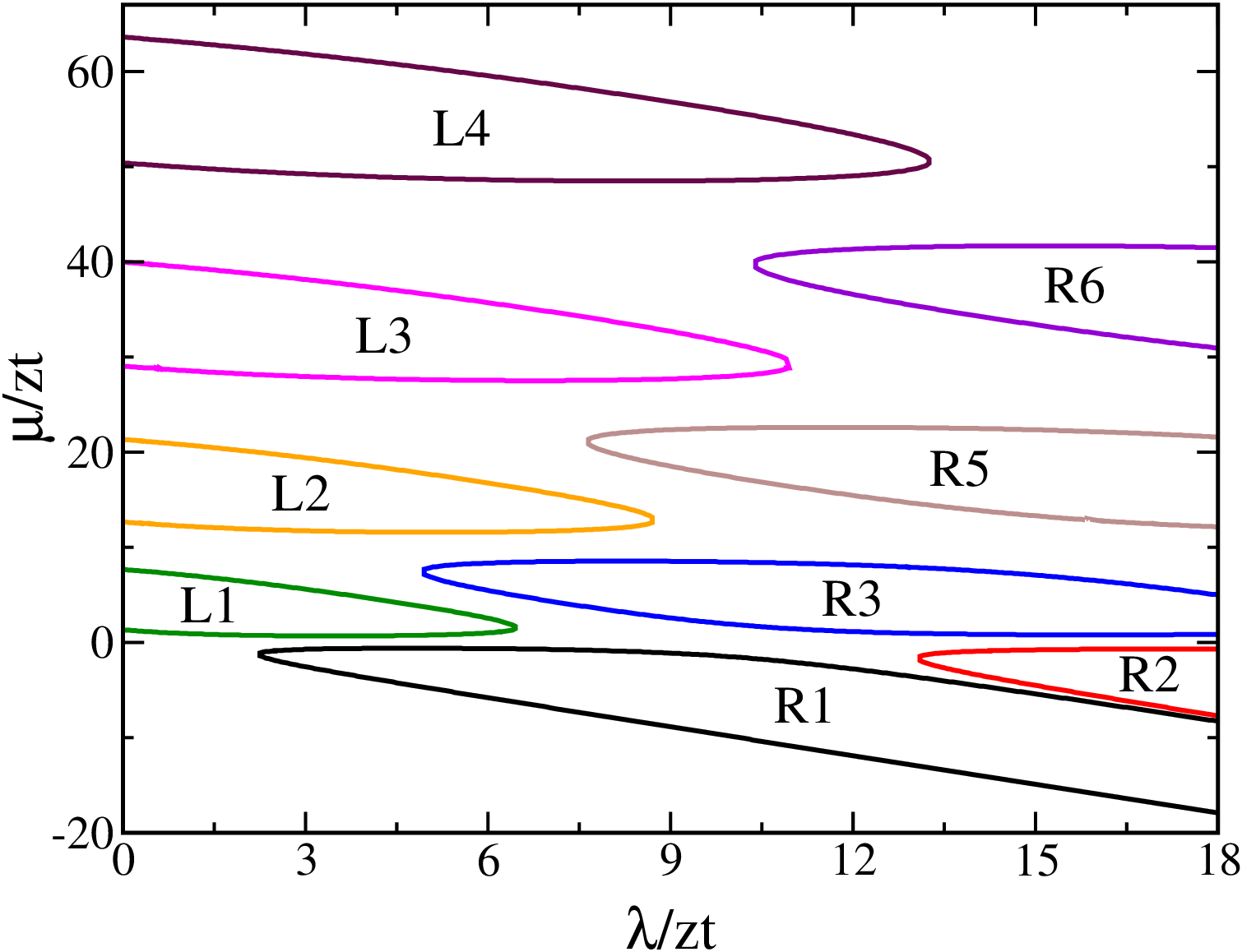, width=6.8cm, height=4.4cm}
\caption{(Color online) $\overline{\lambda}$ - $\overline{\mu}$ phase diagram
for $\overline{U}=10$, $\overline{W}=5.0$, for optical superlattice}
 \label{fig:fig4}
\end{figure}
For an optical superlattice we show the effect of $\overline{W}$ on MI and SLMI phases
in the phase diagram plotted in the $\overline{\mu}$ - $\overline{\lambda}$ plane 
(Figs.~\ref{fig:fig3} and ~\ref{fig:fig4}).
In Fig.~\ref{fig:fig3} we present the phase diagram for $\overline{U}=10$ and 
$\overline{W}=0.0$. Fig.~\ref{fig:fig4} is the phase diagram for $\overline{U}=10$ and 
$\overline{W}=5.0$. Lobes L$\rho$
represent the MI phase with density $\rho$. Lobes R1 to R6 represent
SLMI phases with density in sublattices ${\mathcal A}$ (${\mathcal
B}$) respectively given by $1(0)$, $2(0)$, $2(1)$, $3(1)$, $3(2)$
and  $4(2)$.

The DMRG results obtained for the optical lattice are given in
Figs.~\ref{fig:fig7} and~\ref{fig:fig8}, for $\rho=2$ and $3$
respectively. Figures 9(a) and 9(b) are the phase diagrams for
the optical superlattice for $\rho=3/2$ and $2$ respectively, with two
values of ${W}$($=0.0,5.0$).

\begin{figure}[t]
 \centering
\psfig{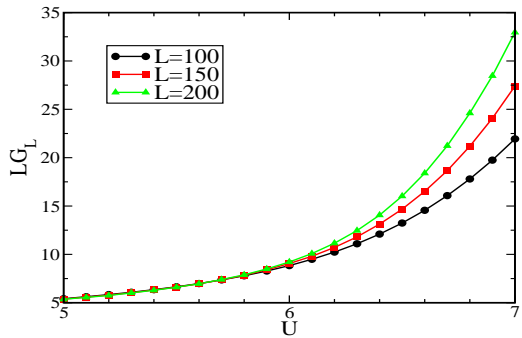} \caption{(Color
online) Scaling of gap $LG_L$ plotted as function of $U$ for
$\rho=2$ and $W=0.0.$} \label{fig:fig5}
\end{figure}

\begin{figure}[b]
 \centering
\psfig{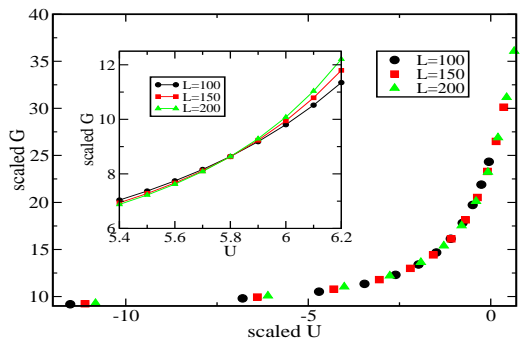}
\caption{(Color online) Scaled gap $G_L$ plotted as a function
of scaled U for $\rho=2$ and $W=0.0$. The curves for different lengths
collapse in the vicinity of $U_C$ as the correlation length $\xi$
diverges exponentially near $U_C$. (Inset) Scaled gap $G_L$ plotted
as a function of U. The curves for different lengths
cross at $U_C$ ($\sim~5.8$) showing the critical point for SF-MI transition.}
\label{fig:fig6}
\end{figure}

For the calculations using the FS-DMRG method, we fix the hopping matrix
element $t=1$ to fix the energy scale (so all the quantities
plotted, are in units of $t$) and to estimate the critical points
$U_C$, we perform finite-size scaling of the single particle gap
$G_L$ (defined by the difference between the energies needed to add
an atom and remove an atom from the system). The plots of $LG_L$ for
different system sizes $L$ (see Fig.(~\ref{fig:fig5})), assuming
that the SF to MI transitions belong to the
Berezinskii-Kosterlitz-Thouless (BKT) type ~\cite{kosterlitz,
giamarchi},
 coalesce in the superfluid phase below the $U_C$.
The value of $U_C$ is then estimated within an error bar of $0.1$ if the values of
 $LG_L$, say for $L=140$ and $200$, differ by less than ${4~\%}$. At the BKT
transition the gap closes satisfying the relation,
${G_L} \sim \exp [-a/|U-U_C|^{1/2}]$, where $a$ is a constant. The correlation length $\xi$,
is finite in the gapped phase and diverges at the critical point as $(1/{G_L})={exp} ~ [a/|U-U_C|^{1/2}]$.
Near $U_C$, the finite-size-scaling relation $LG_{L} [ 1+\{1/({2}~ln~L+C)\}]=F({\xi}/{L})$,
is used to estimate the transition point as done in 
 Ref.~\cite{tapanmrigol}.
Therefore, if we plot $ln(L/\xi)$ vs. $LG_{L} \texttimes \left( 1+(1/({2}~ln~L+C) \right)$
then the curves for different lengths collapse in the vicinity of $U_C$(scaled) (see Fig.(~\ref{fig:fig6}), main panel).
Combining the scaling method described above and DMRG results, we give an approximate value of $U_C$
in various configurations.

\begin{figure}[t]
 \centering
\psfig{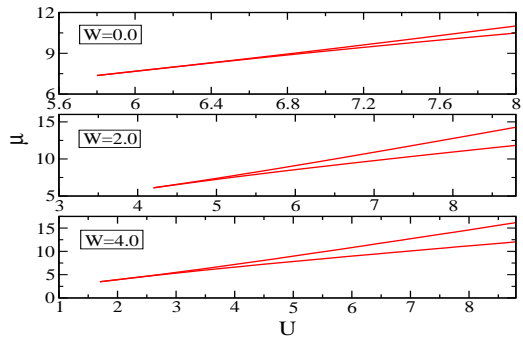}
\caption{(Color online) Phase diagram for $\rho=2$ for various
values of $W$, for optical lattice.} \label{fig:fig7}
\end{figure}

From the mean field results (Figs.~\ref{fig:fig1}
and~\ref{fig:fig2}) for the optical lattice, we find that
the $\rho=1$ MI lobe remains unaltered in the presence of
$\overline{W}$. However, for higher densities, the critical value
$\overline{U}_C(\overline{W})$ for SF-MI transition decreases as
$\overline{W}$ increases (e.g. $\overline{U}_C$ for $\rho=2$ lobe decreases from
$\sim10$ to $\sim7$ when $\overline{W}$ increases from $0.0$ to $4.0$)
and this is more prominent as the density increases, as shown in 
Fig.~\ref{fig:fig1}. Also, the MI
lobes get enlarged as the $\overline{W}$ increases. This is further
confirmed by the DMRG results from Figs.~\ref{fig:fig7}
 and ~\ref{fig:fig8}. The trend is the same for both the densities, $\rho=2$ and $3$, but the effect
of $W$ is more when the density is large. For $\rho=2$,
$U_{C}(W=4.0)\approx 1.6$ compared to $U_{C}(W=0.0)\approx 5.7$. For
$\rho=3$, $U_C$ decreases steadily as $W$ increases; $U_{C}(W=0.0)\approx
8.6$, $U_{C}(W=1.0)\approx6.6$, $U_{C}(W=2.0)\approx4.6$, for
$U_{C}(W=3.0)\approx2.6$ and $U_{C}(W=4.0)\approx0.8$. 
The reason for this behavior at
higher densities is that there is a greater probability of having
three or more atoms at a site, which enhances the three-body
interaction and suppresses atom hopping from one site to another.

\begin{figure}[b]
 \centering
\psfig{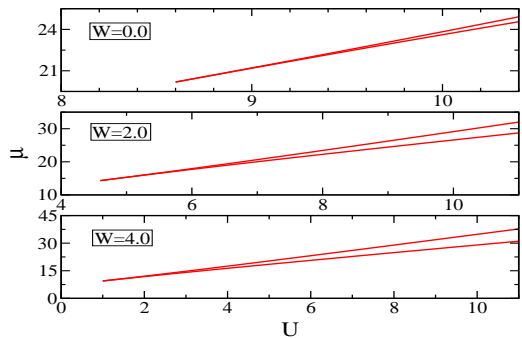}
\caption{(Color online) Phase diagram for $\rho=3$ for various
values of $W$, for optical lattice.} \label{fig:fig8}
\end{figure}

\begin{figure}[t]
 \centering
\psfig{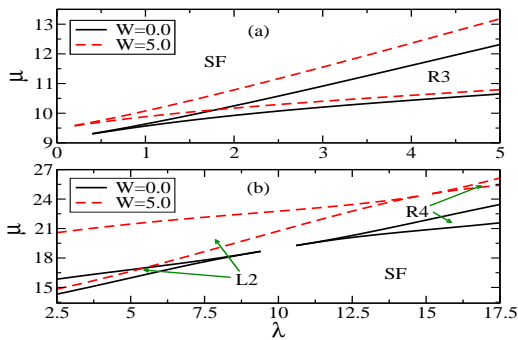}
\caption{(Color online) (a) Phase diagram for $\rho=3/2$ and (b) phase diagram for $\rho=2$, in an optical
superlattice with $W=0.0$ and $5.0$.} \label{Fig.9(a) and 9(b)}
\end{figure}

From the mean field results for the optical superlattice
(Figs.~\ref{fig:fig3} and ~\ref{fig:fig4}),
 we see that the lobes L1, R1 and R2 remain unaffected in the presence
of a finite $\overline{W}$. This is expected because in such
configurations, no two adjacent sites have more than two atoms, and
for an atom to hop two sites is a second order process, which is of
much less probability. However, the SLMI phase R3, which has
sublattice atomic densities $\rho_A=2$ and $\rho_B=1$ (and thus has
average density $\rho=3/2$) gets enlarged in the presence of
$\overline{W}$. This is understood from the following reason. When
$\overline{W}=0.0$ and as we increase $\overline{\lambda}$, keeping the
average density $\rho=3/2$, the ground state goes from the superfluid to
the SLMI phase R3. With a further increase of $\overline{\lambda}> 15$, the
ground state is again a superfluid which has $2 < \rho_A < 3$ and $0
< \rho_B < 1$. However, when $\overline{W}$ is finite, the system does
not prefer to have sublattice densities above 2. Thus SLMI phase R3
has a lower energy than superfluid with sublattice densities $2 <
\rho_A < 3$ and $0 < \rho_B < 1$.
 DMRG results [Fig.9(a)] show a similar trend. When $W=0.0$, the system undergoes a phase
transition from the SF phase to the SLMI phase R3 at a value of $\lambda\sim0.3$. However, in the presence
 of a finite $W$, the transition occurs at a lower
$\lambda(\sim0.15)$, signifying the enlargement of the insulating
lobes.

The phase diagram for the system with a filling factor $\rho=2$, shows
a marked difference in the presence of $\overline{W}$. Comparing
Figs.~\ref{fig:fig3} and~\ref{fig:fig4}, we find that the MI lobe L2
becomes large and its tip shifts from $\overline{\lambda}\sim 1$ to
$\overline{\lambda}\sim 9$. Also as $\overline{\lambda}$ increases
to $\sim17.5$ (Fig.~\ref{fig:fig3}), the MI lobe L2 goes to the SLMI phase R4. 
However, in the presence of $\overline{W}$ (Fig.~\ref{fig:fig4}), the tip of the R4
lobe gets shifted to $\sim 20.5$. As we have considered the maximum value of 
$\overline{\lambda}$ till 18.0, SLMI phase R4 does not appear in
Fig.~\ref{fig:fig4}. Similar results are also obtained by the DMRG
analysis (Fig.9(b)). For finite $W=5.0$, the critical superlattice
potential $\lambda_C$ for transition from the MI phase L2 to SF phase shifts from $9.4$ to $14.8$ and
that for the SF phase to SLMI phase R4 shifts from $10.6$ to $15.2$.
In the absence of $W$ and for lower values of $\lambda$, the
superlattice initially stays in the MI phase L2, for $U=10.0$. As
$\lambda$ becomes comparable to $U$, the system goes from the MI
phase to the SF phase at $\lambda\sim9.4$. As $\lambda$ is further
increased, the system goes from the SF phase to
 the SLMI phase R4, at a value of $\lambda \sim 10.6$.  For $W=5.0$, the system initially is in the MI phase, L2.
Now, due to the presence of $W$, the SF window is shifted to a
$\lambda$ value which is comparable to $U+W$ as shown in
Fig.9(b). The MI to SF transition takes place at
$\lambda\sim14.8$, and the second transition from SF to SLMI
(R4) takes place at $\lambda\sim15.2$. The SF window not only
shifts for $W=5.0$ but also shrinks when compared to that of $W=0.0$. This shifting of the R4 lobe can be understood as follows:
in the R4 phase there are three atoms at every alternate site. As $W$ increases, it becomes difficult to confine three or more
atoms at a single site. Hence to suppress this effect due to the increase in $W$, we need to increase $\lambda$.

The three-body interaction strength scales with the two-body interaction
strength as follows:
$\overline{W} \varpropto \ln({C{\eta}^2}) {(V_0/E_r)^{3/4}}$ $e^{-2\sqrt{V_0/E_r}} {a_{s}^{2}} {k}^2 {\overline{U}}^2$
~\cite{zhang,zhang-jltp,koehler,wu}.
The typical range of  ${a_{s}^{2}} {k}^2$ is $10^{-8}$ to $10^{-2}$ which supports the fact that 
the three-body interaction is weaker than the two-body interaction ~\cite{pethick}. Also it can be
seen that three-body interaction is tunable and can be adjusted by varying $(V_0/E_r)$.
The three-body effects have been experimentally observed before
through various methods as mentioned
earlier~\cite{sewill,nagerl,ruichao}. 
We propose an alternate method
to observe these effects in an optical lattice and superlattice that
we have considered in our present work. The effect of $W$ is very
small compared to the two-body interaction in the system of bosons
in an optical lattice. 
This is of course true when the filling
factor of the system is unity. From Eq(~\ref{eq:one}) it is clear
that the three-body energy scales as $n^3$. Therefore, in order to
observe the effect of the three-body interaction in the experiment it is
important to study the SF-MI transition at higher densities. In the
seminal work of  Greiner \textit{et al.}~\cite{greiner}, the SF-MI transition
was observed by probing the excitation spectrum resulting from a
particle-hole excitation. Such an excitation was created by applying
a potential gradient to the system in the MI phase. By plotting the
excitation probability versus an applied vertical potential
gradient, two narrow resonance peaks were seen. The first peak was
at the potential gradient equal to the single particle excitation
gap, and this corresponds to the MI shell at density equal to one.
One of the possible reasons for the appearance of the second
peak was the particle-hole excitation created in the MI shell at a density
equal to two. In the MI shell at a density equal to two, the
particle-hole excitation at a given site would populate one of the
neighboring sites with three-atoms. In principle, when there are three
or more atoms in a lattice site, the atoms will experience the
effect of $W$ along with that of $U$. In general when there are $n$ ($ >1 $) atoms in each site the system
is in the MI phase with a density equal to $n$,
the excitation gap is $\Delta=U+(n-1)W$ for the optical lattice and
 $\Delta=U+(n-1)W+\lambda$ for the optical superlattice.
Therefore, by measuring the values of the potential gradients for the higher order peaks,
and taking the difference between them for different densities,
it would be possible to determine the value of $W$.

In conclusion, we have studied the effects of the onsite three-body interactions
in a system of neutral bosons in an optical lattice
and superlattice. We first use the mean field theory to understand the behavior of the system and
then confirm the results using the DMRG method. In the optical lattice, and the superlattice
as well, we find that the Mott insulator lobes get enlarged as the
value of $W$ increases. When the density $\rho=1$, the effect of
$W$ is not significant. However, as the density of the system
increases the effect of $W$ becomes significantly large which changes the
SF-MI critical point drastically. We obtain the phase diagrams for different combinations of
densities, strengths of the three-body interaction and the superlattice potential. Finally, we have also
suggested a possible experimental scenario by which it may be possible to observe a signature
of the three-body interaction.

We would like to thank Sebastian Will and Immanuel Bloch for 
valuable discussions and for drawing our attention to the experimental
observations of the three-body interactions. We would also like to
acknowledge CDAC-Garuda for providing us with the computing
facilities and UGC, India (R.V.P.) for support.


\begin{thebibliography}{9}

\bibitem{fisher} M.P.A. Fisher, P.B. Weichmann, G. Grinstein and D.S. Fisher,
Phys. Rev. B {\textbf{40}}, 546 (1989).

\bibitem{jaksch} D. Jaksch, C. Bruder, J. I. Cirac, C. W. Gardiner and P. Zoller, Phys. Rev. Lett. {\textbf{81}}, 3108
(1998).

\bibitem{greiner} M Greiner, O. Mandel, T. Esslinger, T. W. Hansch
and I. Bloch, Nature {\bf 415}, 39 (2002).

\bibitem{lewenstein} M. Lewenstein, A. Sanpera, V. Ahufinger, B. Damski, A. Sen De and U. Sen,
Advances in Physics, Vol. \textbf{56}, 243 (2007).

\bibitem{bloch} I. Bloch, J. Dalibard and W. Zwerger, Rev. Mod. Phys. \textbf{80}, 885
(2008).

\bibitem{ibloch} Simon Folling, Artur Widera, Torben Muller, Fabrice Gerbier,
and Immanuel Bloch, Phys. Rev. Lett. \textbf{97}, 060403  (2006͒).

\bibitem{wketterle} G. K. Campbell, J. Mun, M. Boyd, P. Medley, A. E. Leanhardt,
  L. G. Marcassa, D. E. Pritchard, and W. Ketterle, Science \textbf{313},
 649  (2006͒).

\bibitem{zhang} Bo-lun Chen, Xiao-bin Huang, Su-Peng Kou, Yunbo Zhang, Phys. Rev. A \textbf{78}, 043603 (2008).

\bibitem{tiesinga} P. R. Johnson, E. Tiesinga, J. V. Porto, C. J. Williams, New Journal of Physics \textbf{11}, 093022 (2009).

\bibitem{zhou} Kezhao Zhou, Zhaoxin Liang, Zhidong Zhang, Phys. Rev. A \textbf{82}, 013634 (2010).

\bibitem{aryadmrg} Arya Dhar, Tapan Mishra, R. V. Pai, B. P. Das, Phys. Rev. A \textbf{83}, 053621 (2011).

\bibitem{aryamft} Arya Dhar, Manpreet Singh, R. V. Pai, B. P. Das, Phys. Rev. A \textbf{84}, 033631 (2011).

\bibitem{sewill} Sebastian Will, Thorsten Best, Ulrich Schneider, Lucia Hackermüller, Dirk-Sören Lühmann and Immanuel Bloch, Nature \textbf{465}, 197 (2010).

\bibitem{nagerl} M. J. Mark, E. Haller, K. Lauber, J. G. Danzl, A. J. Daley, and H.-C. N\"{a}gerl, Phys. Rev. Lett. \textbf{107}, 175301 (2011).

\bibitem{ruichao} Ruichao Ma, M. Eric tai, Philipp M. Preiss, Waseem S. Bakr, Jonathan Simon, and Markus Greiner, Phys. Rev. Lett. \textbf{107}, 095301 (2011).

\bibitem{stoof} D. van Oosten, P. van der Straten, and H. T. C. Stoof, Phys. Rev. A \textbf{63}, 053601 (2001).

\bibitem{sheshadri} K. Sheshadri, H. R. Krishnamurthy, R. Pandit, and T. V. Ramakrishnan, Europhys. Lett. \textbf{22}, 257
(1993).

\bibitem{kosterlitz} J. M. Kosterlitz and D. J. Thouless, J. Phys. C \textbf{6}, 1181 (1973͒).

\bibitem{giamarchi} Thierry Giamarchi, \textit{Quantum Physics in One Dimension} ͑(Clarendon Press, Oxford, 2004͒).

\bibitem{tapanmrigol} Tapan Mishra, Juan Carrasquilla, and Marcos Rigol, Phys. Rev. B \textbf{84}, 115135 (2011).

\bibitem{zhang-jltp} Y. Zhang and F. Han, Journal of Low Temperature Physics, \textbf{141}, 314 (2005).

\bibitem{koehler} Thorsten K\"{o}hler, Phys. Rev. Lett. \textbf{89}, 210404 (2002).

\bibitem{wu} T. T. Wu, Phys. Rev., \textbf{115}, 1390 (1959).

\bibitem{pethick} C. J. Pethick and H. Smith, \textit{Bose-Einstein Condensation in Dilute Gases},
(Cambridge University Press, Cambridge, England, 2002).

\end{thebibliography}
\end{document}